\documentclass[letterpaper]{article}
\usepackage[preprint]{aaai2027}
\usepackage[hyphens]{url}
\usepackage{graphicx}
\usepackage{natbib}
\usepackage{caption}
\usepackage{booktabs}
\usepackage{amsmath}
\usepackage{amssymb}
\title{SingDance: Compositional Zero-Shot Singing-and-Dancing Video Generation with Role-Aware Audio Conditioning}
\author{
    Tao Feng\textsuperscript{\rm 1,\rm 2}\protect\thanks{Work done during internship at Kling Team, Kuaishou Technology.},
    Xu Li\textsuperscript{\rm 1}\corresponding,
    Xiangyang Luo\textsuperscript{\rm 3},
    Ming Wen\textsuperscript{\rm 1},\\
    Huadai Liu\textsuperscript{\rm 1,\rm 2},
    Chen Zhang\textsuperscript{\rm 1},
    Wei Xue\textsuperscript{\rm 2}\corresponding
}
\affiliations{
    \textsuperscript{\rm 1}Kling Team, Kuaishou Technology\\
    \textsuperscript{\rm 2}The Hong Kong University of Science and Technology\\
    \textsuperscript{\rm 3}Tsinghua University\\
    \{tao.feng,huadai.liu\}@connect.ust.hk, \{lixu15,wenming,zhangchen03\}@kuaishou.com\\
    goodluoxy@gmail.com, weixue@ust.hk
}

\begin{document}
\maketitle

\begin{abstract}
Generating personalized dance videos from a reference image, text prompt, and audio track requires music-conditioned body motion. Singing-and-dancing adds a second requirement: the visible subject must also articulate the vocals. Existing music-conditioned methods focus primarily on choreography, while speech-driven models generally assume that the visible subject produces the input voice, leaving this combined setting largely underexplored. We introduce SingDance, a unified video diffusion framework that formulates controllable vocal articulation as a semantic role: the visible subject is either the \emph{source}, who produces the vocal signal, or the \emph{listener}, who receives it from an off-screen performer. Hard-compact routing selects task-relevant speech, music, and role conditions, which are composed through frame-wise joint audio injection; source and listener retain the same speech pathway. Training uses asymmetric supervision: on-screen speaking and curated off-screen conversational-response videos establish role control, while instrumental and song-based dancing-only videos establish music-conditioned body motion. The target Song/Source configuration is never observed during training. At inference, assigning the source role to a song composes separately learned articulation and song-conditioned dance capabilities, enabling compositional zero-shot singing-and-dancing. Experiments demonstrate strong motion--beat alignment and visual fidelity, reliable paired switching of vocal articulation while preserving music-aligned body motion, and highly competitive lip synchronization with substantially fewer generation-time parameters than the strongest speech-driven baseline evaluated.
\end{abstract}

\begin{links}
    \link{Project Page}{https://fff-ttt.github.io/singdance-project-page/}
\end{links}

\begin{figure}[t]
    \centering
    \includegraphics[width=\columnwidth]{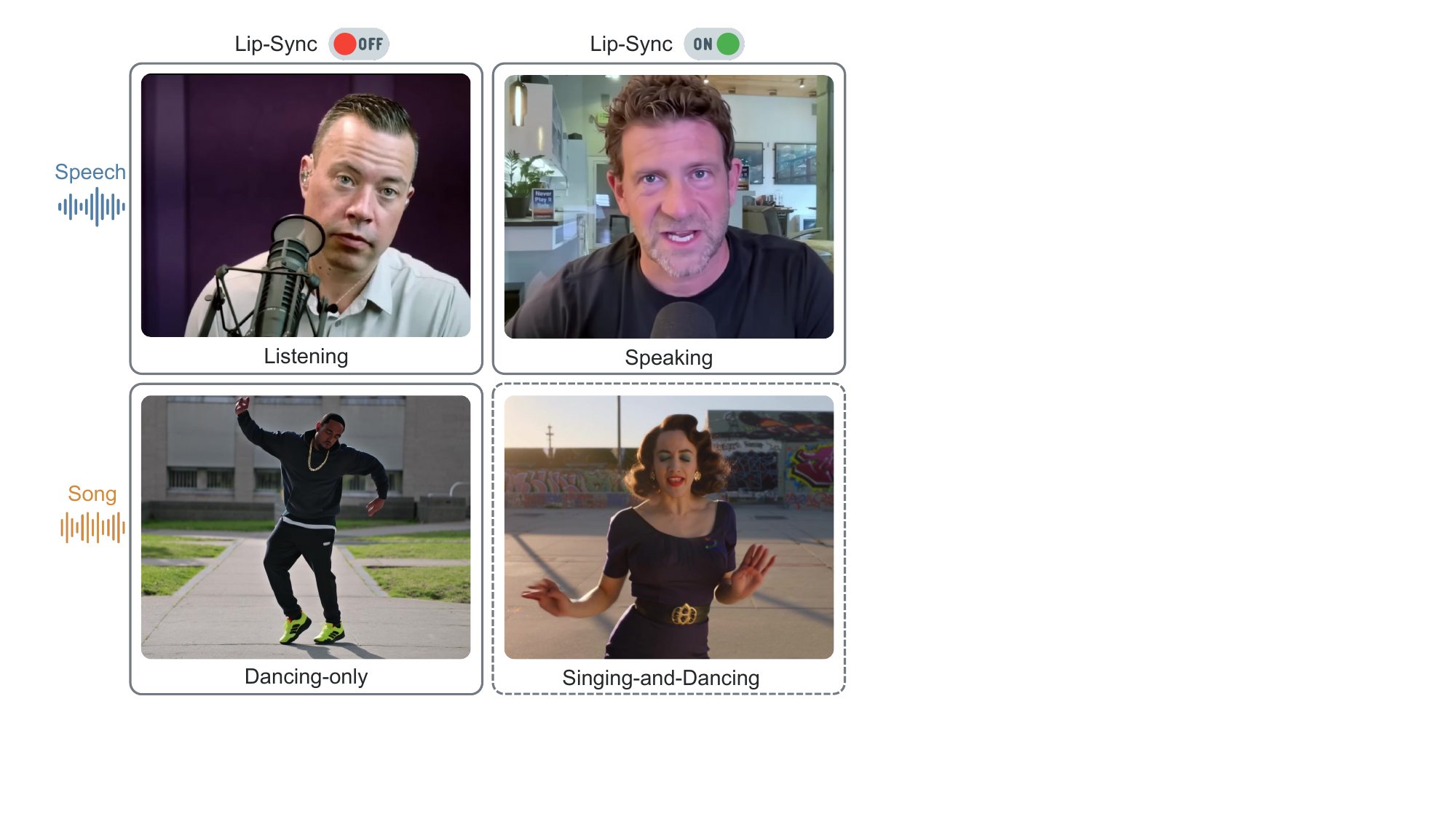}
    \caption{Four generation tasks supported by SingDance. The dashed box marks singing-and-dancing, the held-out Song/Source configuration generated through compositional zero-shot inference.}
    \label{fig:teaser}
\end{figure}

\section{Introduction}

Short-form dance videos are a major form of online visual content. They broadly take two forms: \emph{dancing-only}, where an on-screen subject moves to instrumental music, or to a song without lip-syncing to its vocals, and \emph{singing-and-dancing}, where dance motion and vocal articulation occur together. Given the same subject and song, creators may wish to produce either form. A single framework supporting both from a reference image, text prompt, and audio track would substantially simplify personalized video creation. Despite its practical relevance, singing-and-dancing remains largely unexplored in music-conditioned human video generation.

Existing research addresses adjacent parts of this problem under different assumptions. Music-conditioned dance methods focus primarily on body choreography and rhythmic alignment, and generally do not model vocal articulation by the dancer~\cite{chen2025xdancer,hong2026musicinfuser,yang2026omnidance,huang2026wandancer}. Speech-driven human video models achieve accurate articulation and expressive motion, but conventionally assume that the input voice is produced by the visible subject~\cite{gao2025wans2v,gan2025omniavatar,yang2025infinitetalk,meng2025echomimicv3}. Recent interactive avatar systems further model speaking and listening as distinct conversational phases, often with separated or phase-specific audio pathways~\cite{zeng2026lpm,sun2026streamavatar}. Song-conditioned dance raises a different question: how to control whether the visible subject produces the vocals while retaining the same vocal content as part of the conditioning for body motion.

Our key insight is to model controllable vocal articulation as an explicit semantic role rather than an isolated mouth-motion instruction. We define the visible subject as either the \emph{source}, who produces the vocal signal, or the \emph{listener}, who receives it from an off-screen performer. This distinction prevents the presence of vocal content from automatically assigning articulation to the visible subject. Genuine listening/reacting examples provide non-speaking supervision for the listener role, rather than defining Lip-Sync OFF as a frozen-mouth instruction. This semantic distinction transfers naturally from speech to songs, organizing four behaviors---speaking, listening, dancing-only, and singing-and-dancing---within a single generation problem.

We instantiate this formulation in SingDance, a video diffusion transformer conditioned on a reference image, text, representations specialized for speech and music, and an explicit vocal role. Hard-compact routing retains the task-relevant token groups: speech interaction uses speech and role tokens, instrumental dancing-only uses music and role tokens, and song-based dancing-only uses all three. Importantly, the two vocal roles share the same speech pathway; compact role conditions work with the prompt's \emph{vocal role} field to specify whether the visible subject is assigned the vocals. For song-based dancing-only, vocal features remain available because lyrics and vocal phrasing may still inform the performance even when they are not articulated by the visible subject.

Training follows a staged design with asymmetric supervision. We first learn role control from on-screen speaking clips and carefully curated conversational clips in which the visible subject genuinely listens or reacts to an off-screen interlocutor. We then learn music-conditioned dance from instrumental and song-based dancing-only videos, without singing-and-dancing supervision. At inference, singing-and-dancing reuses the complete speech-plus-music configuration learned from song-based dancing-only and changes only the vocal-role state from listener to source. This state is expressed consistently through the learned role conditions and the prompt's \emph{vocal role} field. This compact intervention enables paired role switching and instantiates the held-out Song/Source configuration through compositional zero-shot inference.

Our contributions are threefold:
\begin{itemize}
    \item We formulate reference-image-conditioned dance video generation with an explicit vocal role, organizing speaking, listening, dancing-only, and singing-and-dancing as four related generation behaviors.
    \item We introduce a role-aware conditioning design in which source and listener share the same speech representation and injection pathway, while task-routed speech, music, and role tokens are composed through frame-wise joint audio injection.
    \item We learn source-role articulation and song-conditioned dance from asymmetric supervision and demonstrate their compositional zero-shot generalization to the held-out Song/Source configuration. Paired role intervention verifies controllable articulation, while motion--beat evaluation and music-conditioning ablation validate music-responsive body motion.
\end{itemize}

\section{Related Work}

\begin{figure*}[t]
    \centering
    \includegraphics[width=\textwidth]{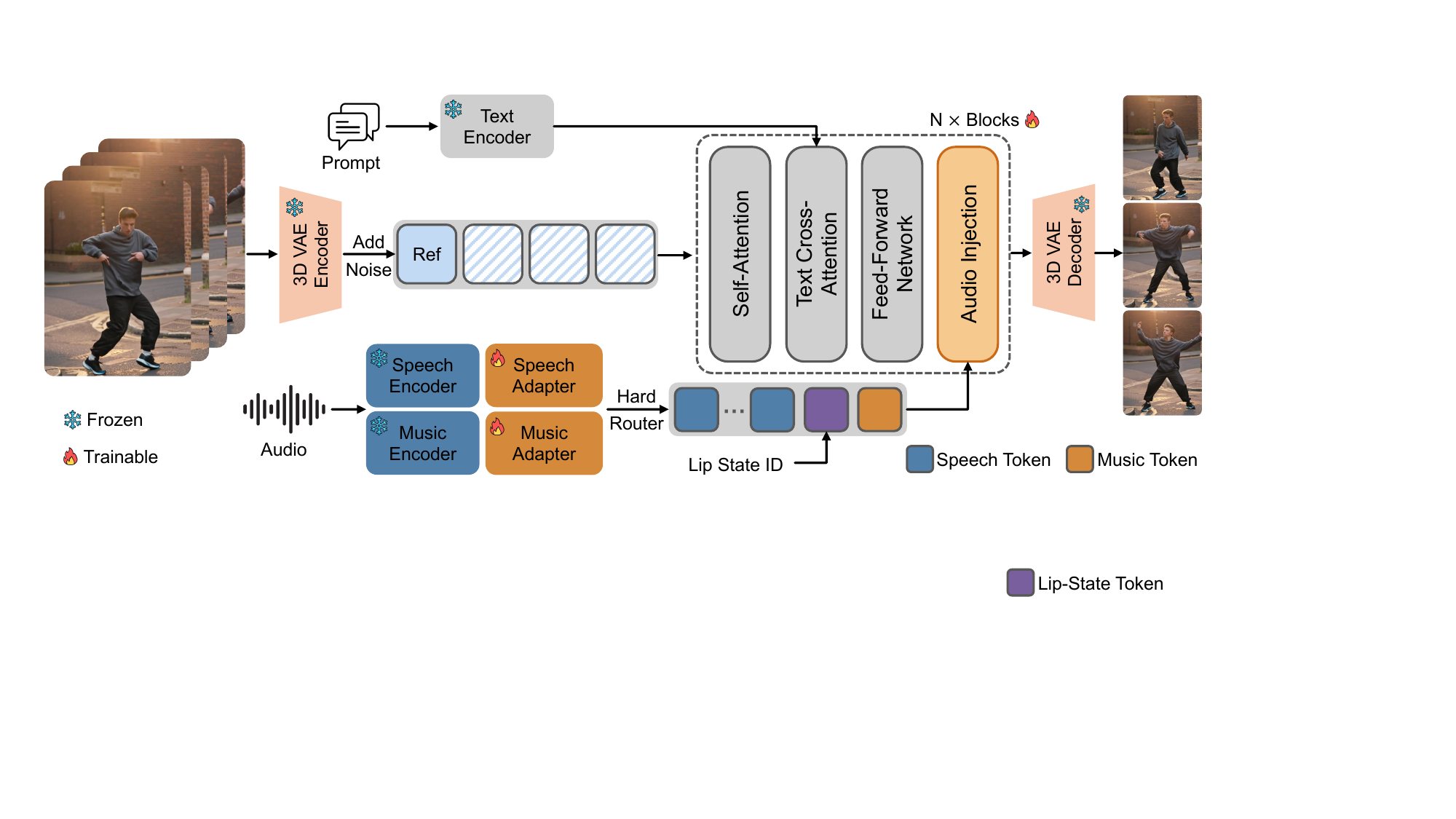}
    \caption{Training architecture of SingDance. The first video frame is separately encoded as a clean reference latent, while subsequent frames are temporally compressed and noised in latent space. Wav2Vec~2.0 and MuQ features, together with the vocal-role condition, are selected by hard-compact routing and supplied to frame-wise joint audio injection. The repeated block is schematic: joint audio injection is applied immediately after each of 10 selected blocks in the 30-block DiT backbone. The denoised latents are decoded into output frames. Snowflakes and flames denote frozen and trainable modules, respectively.}
    \label{fig:method-overview}
\end{figure*}

\paragraph{Music-conditioned dance generation.}
Music-conditioned dance generation is inherently one-to-many, and prior work adopts different intermediate representations to structure this ambiguity. Early methods generate 3D body motion before rendering, including FACT on the AIST++ dataset~\cite{li2021aist}, Bailando~\cite{siyao2022bailando}, EDGE~\cite{tseng2022edge}, and FineDance~\cite{li2023finedance}. More recent systems move toward RGB video while retaining explicit motion intermediates: X-Dancer predicts music-aligned 2D pose tokens~\cite{chen2025xdancer}, whereas MACE-Dance cascades motion and appearance experts~\cite{yang2025mace}. A complementary line directly adapts video diffusion models. MusicInfuser introduces music conditioning into a text-to-video backbone~\cite{hong2026musicinfuser}. Concurrent studies further explore different allocations between text and music: OmniDance combines image, text, and MERT features in a 5B video backbone~\cite{yang2026omnidance}, while the more recent Wan-Dancer uses coarse textual attributes and hierarchical global-to-local music conditioning for long-form generation~\cite{huang2026wandancer}. These concurrent methods explore different allocations of text and music for choreographic control. SingDance studies an additional compositional axis: whether the visible dancer produces the vocals while retaining music-conditioned body motion.

\paragraph{Speech-driven human video generation.}
Speech-driven human video generation spans subject-specific neural rendering, including audio-driven 3D Gaussian talking heads~\cite{xie2025pointtalk}, and generalizable diffusion-based models that have substantially improved identity preservation, lip synchronization, and natural co-speech motion~\cite{chen2025hunyuan,gan2025omniavatar,gao2025wans2v,yang2025infinitetalk,meng2025echomimicv3,longcat2026avatar}. Complementary work studies general audio--visual synchronization representations for evaluation and generation guidance~\cite{feng2025unisync}. Most systems, however, assume that the visible subject produces the input utterance and rely on speech-pretrained encoders, such as Wav2Vec or Whisper, for temporally aligned conditioning~\cite{baevski2020wav2vec,radford2023whisper}. Although these representations capture phonetic and prosodic cues for articulation, they are not explicitly optimized for music-specific structures such as beat, timbre, and longer-range rhythm. Interactive-avatar and 3D facial-animation systems instead model speaking and listening through dual-speaker context, listener-specific motion priors, or phase-specific audio pathways~\cite{peng2025dualtalk,chu2026unils,wang2026flowact,weng2026beyondmonologue,zhang2026embodiedhead}. DualTalk models two-speaker interaction~\cite{peng2025dualtalk}; UniLS separates an internal listener-motion prior from external audio modulation~\cite{chu2026unils}; LPM~1.0 uses independently parameterized speaking and listening streams~\cite{zeng2026lpm}; and StreamAvatar routes speech features through dedicated Talk and Listen attention modules~\cite{sun2026streamavatar}. These designs account for distinct conversational motion distributions. SingDance instead keeps the speech representation and conditioning pathway shared across roles, changing coordinated textual and learned role conditions. This permits held-out Song/Source composition while retaining the speech and music conditions used for song-based dancing-only.

\paragraph{Multimodal conditioning in video foundation models.}
Pretrained video diffusion models provide reusable foundations that can be extended with additional control modalities. Wan-S2V combines text-based global motion specification with frame-aligned speech injection~\cite{gao2025wans2v}. ActAvatar builds upon Wan2.2-TI2V-5B and combines temporally structured text conditioning for phase-level actions with Wav2Vec~2.0-based speech cross-attention for articulation~\cite{peng2026actavatar}. For music-conditioned generation, OmniDance augments the same TI2V backbone with MERT-based music cross-attention, whereas Wan-Dancer adapts Wan-I2V using compact Librosa features in a hierarchical generation pipeline~\cite{yang2026omnidance,huang2026wandancer}. These approaches illustrate different ways of assigning control responsibilities to text and audio while preserving the generative prior of a pretrained video model. SingDance assigns each modality a primary responsibility according to the information it represents most naturally: the reference image anchors appearance and the initial scene; text specifies intended action and camera evolution; speech captures vocal articulation and phrasing cues; and music complements the prompt with rhythm and broader musical structure that are difficult to express textually. This allocation motivates separate speech and music representations, composed through frame-wise joint audio injection, while explicit vocal-role control determines whether the visible subject articulates the vocals.

\section{Method}

In this section, we first introduce the audio representations and vocal-role control, then describe hard-compact routing and joint audio injection, and finally present staged training for zero-shot composition and text--audio classifier-free guidance.

\subsection{Audio Representations and Vocal-Role Control}

To coordinate vocal articulation with music-conditioned body motion within a single generation framework, SingDance employs representations specialized for speech and music together with coordinated textual and learned vocal-role controls.

As illustrated in Figure~\ref{fig:method-overview}, we use Wav2Vec~2.0~\cite{baevski2020wav2vec} to extract speech-oriented features for vocal articulation. Following Wan-S2V~\cite{gao2025wans2v}, we aggregate its layer-wise representations using learnable weights and map them through a temporal encoder to a global speech feature and frame-aligned local speech tokens. In parallel, MuQ~\cite{zhu2025muq} provides music-oriented acoustic and structural features, which a separate temporal encoder maps to frame-aligned local music tokens for body-motion conditioning. Both temporal encoders use non-causal 1D convolutions, matching the non-causal context of the pretrained features.

At the level of generated behavior, the two vocal roles correspond to the Lip-Sync ON and OFF states shown in Figure~\ref{fig:teaser}. Semantically, \emph{source} indicates that the vocal signal is produced by the visible subject, whereas \emph{listener} indicates that it originates from an external source. We thus treat lip synchronization as an outcome of vocal-role assignment rather than an isolated mouth-motion instruction.

To expose this distinction to the text encoder while separating vocal behavior from choreography, we organize each prompt into five fields: \emph{vocal role}, \emph{style}, \emph{action}, \emph{first frame}, and \emph{camera}. The standardized \emph{vocal role} field assigns the vocal signal either to the visible subject or to an external source, corresponding to \emph{source} for Lip-Sync ON and \emph{listener} for Lip-Sync OFF. During prompt construction, we remove references to lip movement, speaking, and singing from the \emph{action} field, retaining only body-motion descriptions. Within the text prompt, vocal behavior is therefore specified exclusively by the \emph{vocal role} field, separating its textual control from the intended body-action sequence.

In parallel, the same binary source/listener state is encoded by a learned global role embedding and a local role token. Each textual role assignment is paired consistently with its corresponding learned role conditions throughout training and inference. Speech and vocal role therefore provide both global and local conditions, whereas music is introduced only through local tokens. When switching roles, SingDance changes only the vocal-role state; its textual and learned representations are updated together, while the role-independent prompt content, speech and music representations, and shared conditioning pathway remain unchanged. This coordinated but compact intervention provides the basis for the held-out composition described later.

\subsection{Hard-Compact Routing and Joint Audio Injection}

Different generation tasks require different subsets of the available conditions. We therefore use \emph{hard-compact routing}: the route is specified by the generation task rather than inferred from audio content, and only the task-relevant speech, music, and role tokens are retained as the routed tokens. Table~\ref{tab:routes} summarizes the five configurations.

\begin{table}[!t]
\centering
{\small
\setlength{\tabcolsep}{2pt}
\begin{tabular*}{\columnwidth}{@{\extracolsep{\fill}}llcl@{}}
\toprule
Audio type & Role & Routed tokens & Task \\
\midrule
Speech & Source & $S+R_{\mathrm{src}}$ & Speaking \\
Speech & Listener & $S+R_{\mathrm{lis}}$ & Listening/Reacting \\
Instrumental & Listener & $R_{\mathrm{lis}}+M$ & Dancing-Only \\
Song & Listener & $S+R_{\mathrm{lis}}+M$ & Dancing-Only \\
\midrule
Song & Source & $S+R_{\mathrm{src}}+M$ & Singing-and-Dancing \\
\bottomrule
\end{tabular*}
}
\caption{Hard-compact routing across generation tasks. $S$, $M$, and $R$ denote speech, music, and role tokens, respectively. The first four routes are used during training; Song/Source is activated only at inference as the held-out compositional configuration.}
\label{tab:routes}
\end{table}

Speech examples retain speech and role tokens. Instrumental dance retains music and role tokens, making music the only time-varying audio input and providing a direct learning signal to the newly introduced music pathway. Song-based dance retains all three token groups under the listener role, because lyrics, vocal phrasing, and phrase boundaries may still inform choreographic timing even when the visible subject does not articulate the vocals. At inference, the held-out Song/Source configuration preserves the speech and music tokens and changes only the vocal-role state, updating its learned role conditions and the prompt's \emph{vocal role} field consistently.

\begin{figure}[t]
    \centering
    \includegraphics[width=.95\linewidth]{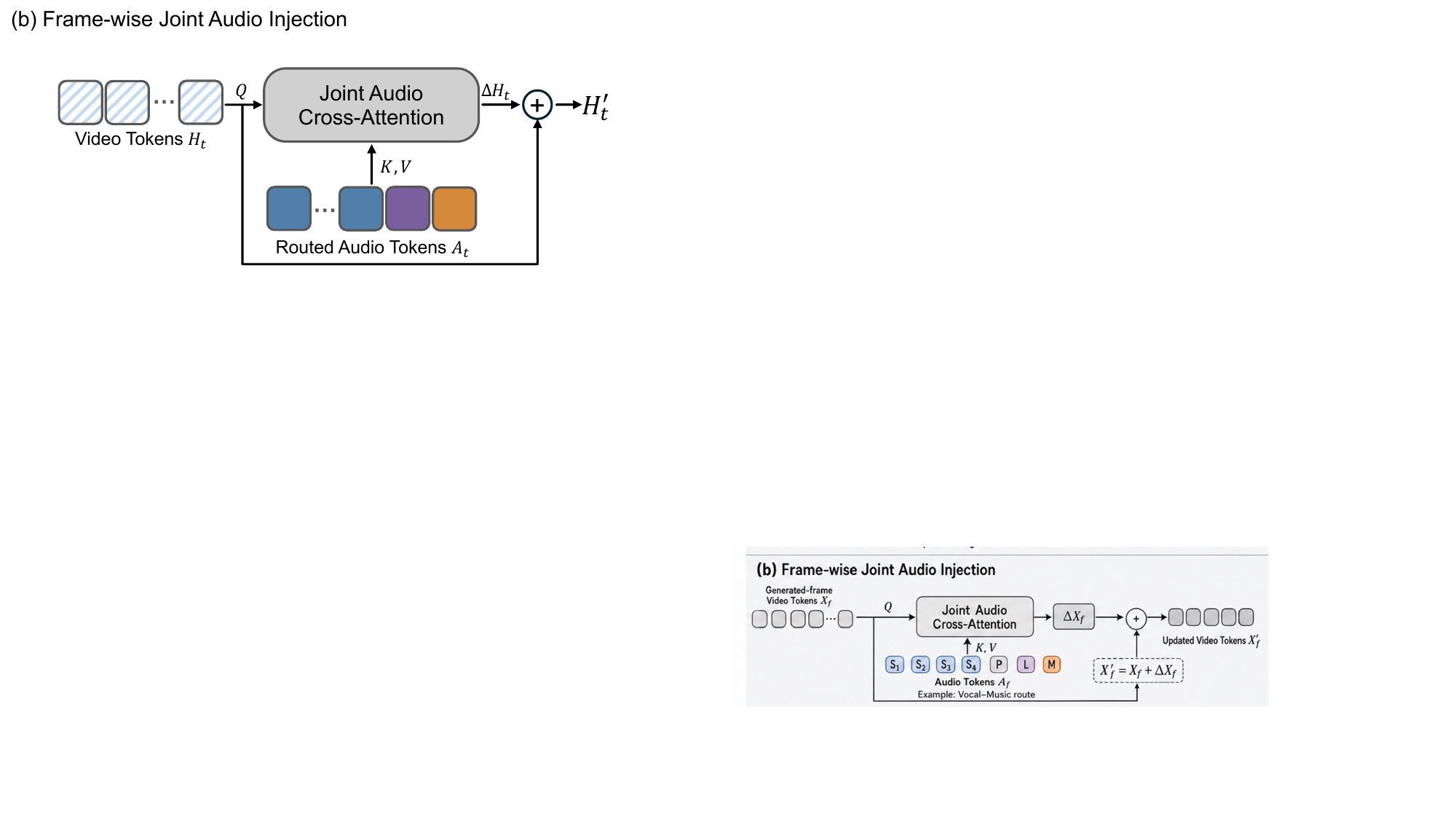}
    \caption{Frame-wise joint audio injection. For each generated latent frame, visual tokens act as queries and the temporally aligned routed tokens serve as joint keys and values; the cross-attention output is added residually to update the visual representation.}
    \label{fig:audio-injection}
\end{figure}

Figure~\ref{fig:audio-injection} details frame-wise joint audio injection, where the routed tokens form a joint key--value sequence temporally aligned with each generated latent frame. For global modulation, speech and song examples use the global speech feature together with the global role embedding, whereas instrumental examples use the global role embedding alone; music is supplied through the routed local tokens. Text remains in the backbone's original text cross-attention, and joint audio injection updates only generated-frame tokens, leaving the separately encoded reference latent unchanged.

\subsection{Staged Training for Zero-Shot Composition}

SingDance adopts a staged training strategy that first establishes vocal-role control and then introduces music-conditioned body motion. This organization reflects the asymmetric supervision available to the two capabilities: speech videos provide source/listener supervision, whereas dancing-only videos provide music-conditioned body-motion supervision. Importantly, no paired singing-and-dancing videos are used in either stage.

In the speech stage, the music pathway is absent. On-screen vocal production grounds the \emph{source} role, while genuine listening and reacting behavior toward an off-screen interlocutor grounds the \emph{listener} role. Both roles share the same speech representation and conditioning pathway, with their assignment expressed through the prompt's \emph{vocal role} field and global/local role conditions. The model therefore learns role-dependent interpretations of a shared vocal signal before music conditioning is introduced.

The dance stage starts from the speech-stage model and introduces the music pathway using instrumental and song-based dancing-only videos. Instrumental examples retain only music and role conditions, making music the sole time-varying audio input. Song examples retain both speech and music conditions because vocal phrasing and phrase boundaries may remain relevant to choreography even when the visible subject does not articulate the vocals. During this stage, speech examples from both roles constitute 20\% of the training samples to mitigate forgetting of articulation and role control.

Across the two stages, SingDance observes four configurations: Speech/Source, Speech/Listener, Instrumental/Listener, and Song/Listener. The target Song/Source configuration is never observed during training. At inference, we instantiate it by changing the \emph{vocal role} field from an off-screen to an on-screen source assignment and switching the corresponding global and local role conditions from listener to source. The reference image, soundtrack, remaining prompt fields, and extracted speech and music features remain unchanged. This composition combines the source-role articulation learned from speech with the joint speech--music conditioning learned from dancing-only videos, without introducing a new generation module or a post-hoc lip-synchronization model.

Both stages use the backbone's standard flow-matching objective~\cite{lipman2023flow} over generated latent positions, without specialized lip-synchronization or beat-alignment losses.

\subsection{Text--Audio Classifier-Free Guidance}

At inference, we use three-pass classifier-free guidance (CFG)~\cite{ho2022classifierfree} with separate text and audio guidance scales. Let $\mathbf{v}_{\bar{t},0}$, $\mathbf{v}_{t,0}$, and $\mathbf{v}_{t,a}$ denote the predictions obtained with the negative prompt and null audio, the input prompt and null audio, and the input prompt with routed audio, respectively. We compute
\begin{equation}
    \widehat{\mathbf{v}}
    = \mathbf{v}_{\bar{t},0}
    + s_t\left(\mathbf{v}_{t,0}-\mathbf{v}_{\bar{t},0}\right)
    + s_a\left(\mathbf{v}_{t,a}-\mathbf{v}_{t,0}\right),
\label{eq:text-audio-cfg}
\end{equation}
where $s_t$ and $s_a$ are the text and audio guidance scales, respectively. The learned vocal-role embeddings remain active in all three passes. The textual \emph{vocal role} field is part of the input prompt and is therefore included in $\mathbf{v}_{t,0}$ and $\mathbf{v}_{t,a}$.

To enable this three-pass CFG, we apply condition dropout during training: text alone is dropped with probability 10\%, audio alone with probability 10\%, and both are dropped with probability 5\%; all remaining samples retain both conditions. Text dropout removes the entire prompt, including its \emph{vocal role} field, whereas the learned vocal-role embeddings are retained in all dropout cases.

\begin{table*}[!t]
\centering
{\small
\setlength{\tabcolsep}{2.6pt}
\begin{tabular*}{\textwidth}{@{\extracolsep{\fill}}llccccccccc@{}}
\toprule
& & \multicolumn{2}{c}{Reference Fidelity} & \multicolumn{2}{c}{Motion--Beat Alignment} & \multicolumn{3}{c}{Visual Quality} & \multicolumn{2}{c}{Human Preference} \\
\cmidrule(lr){3-4}\cmidrule(lr){5-6}\cmidrule(lr){7-9}\cmidrule(lr){10-11}
Method & Type & Subject $\uparrow$ & Background $\uparrow$ & BeatAlign $\uparrow$ & MBCR $\uparrow$ & Aesthetic $\uparrow$ & Imaging $\uparrow$ & Smoothness $\uparrow$ & Rhythm $\uparrow$ & Visual $\uparrow$ \\
\midrule
\multicolumn{11}{@{}l}{\textbf{Singing-and-Dancing}} \\
GT & -- & 0.9673 & 0.9735 & 0.2795 & 0.2922 & 0.5720 & 0.6553 & 0.9858 & -- & -- \\
MusicInfuser$^\dagger$ & TA2V & -- & -- & 0.2356 & \underline{0.2337} & 0.5232 & 0.6172 & \textbf{0.9927} & -- & -- \\
Wan-S2V & TIA2V & \underline{0.9565} & \underline{0.9633} & \underline{0.2640} & 0.1983 & \underline{0.5834} & \underline{0.6538} & 0.9860 & \underline{38.67} & \underline{45.33} \\
\textbf{SingDance} & TIA2V & \textbf{0.9682} & \textbf{0.9772} & \textbf{0.2723} & \textbf{0.2558} & \textbf{0.5846} & \textbf{0.6688} & \underline{0.9899} & \textbf{61.33} & \textbf{54.67} \\
\midrule
\multicolumn{11}{@{}l}{\textbf{Dancing-Only}} \\
GT & -- & 0.9648 & 0.9705 & 0.3051 & 0.3027 & 0.5738 & 0.6667 & 0.9880 & -- & -- \\
MusicInfuser$^\dagger$ & TA2V & -- & -- & \underline{0.2872} & \textbf{0.2661} & 0.5258 & 0.6522 & \textbf{0.9948} & -- & -- \\
Wan-S2V & TIA2V & \underline{0.9415} & \underline{0.9493} & 0.2467 & 0.2180 & \textbf{0.5776} & \underline{0.6570} & 0.9867 & \underline{32.00} & \underline{44.33} \\
\textbf{SingDance} & TIA2V & \textbf{0.9705} & \textbf{0.9766} & \textbf{0.3002} & \underline{0.2632} & \underline{0.5718} & \textbf{0.6738} & \underline{0.9921} & \textbf{68.00} & \textbf{55.67} \\
\bottomrule
\end{tabular*}
}
\caption{Quantitative results on \textit{SingDance-50} and \textit{Dance-100}. Human scores are preference rates (\%). Best and second-best generated results are bold and underlined; GT is shown only as a reference. $^\dagger$ MusicInfuser does not accept a reference image, so its reference-fidelity scores are not applicable.}
\label{tab:main-dance}
\end{table*}

\section{Experiments}

\subsection{Experimental Setup}

\paragraph{Data.}
The speech phase combines HuMoSet~\cite{chen2025humo} with a proprietary web-video collection, totaling approximately 300,000 five-second clips. Among them, approximately 12,000 clips depict genuine listening or reacting behavior toward an off-screen interlocutor, while the remainder primarily depict on-screen speaking. The dance phase combines cleaned MA-Data~\cite{yang2025mace} with a proprietary dance-video collection, yielding approximately 48,000 five-second clips: 18,000 with instrumental music and 30,000 with songs. A small fraction of song-conditioned clips exhibits detectable or uncertain lip synchronization; we remove these cases using SyncNet~\cite{chung2017out} and retain the remainder as dancing-only supervision. No singing-and-dancing videos are used for training.

For dance evaluation, we curate two test sets disjoint from all training data: \textit{SingDance-50}, comprising 50 face-visible vocal-song cases for singing-and-dancing, and \textit{Dance-100}, comprising 100 diverse dancing-only cases. For talking, we use EMTD~\cite{meng2024echomimicv2}. The vocal-role intervention reuses \textit{SingDance-50} and changes only the vocal-role assignment, while keeping the reference image, soundtrack, role-independent prompt fields, sampling seed, model checkpoint, and all other inference settings fixed.

\paragraph{Baselines.}
Open-source systems that directly generate RGB dance videos from audio remain limited, as much of the music-to-dance literature generates intermediate motion representations. We therefore select two complementary end-to-end baselines. MusicInfuser~\cite{hong2026musicinfuser} directly conditions a text-to-video model on music; its text--audio-to-video (TA2V) formulation offers a representative comparison for music-responsive generation even without reference-image conditioning or explicit vocal articulation. Wan-S2V~\cite{gao2025wans2v} accepts text, image, and audio inputs (TIA2V), shares the Wan video-model family with SingDance, and provides a particularly relevant speech-conditioned human-animation baseline; our speech pathway is also inspired by its frame-wise audio injection. Its formulation assigns the input vocals to the visible subject and uses speech-oriented rather than music-specific audio conditioning. Together, they provide references for music-conditioned dancing and speech-conditioned human animation. We evaluate both on both dance test sets using their supported input interfaces and recommended prompt formats.

\paragraph{Implementation Details.}
All SingDance results are produced by a single model initialized from Wan2.2-TI2V-5B~\cite{wan2025}. We train on 64 NVIDIA A100 GPUs with an effective batch size of 64, using AdamW~\cite{loshchilov2019decoupled} with a learning rate of $1\times10^{-5}$. The speech and dance phases are each trained for eight epochs at 480p, after which the dance phase is further fine-tuned for four epochs, approximately 3K optimization steps, at $704\times1280$. At inference, talking videos follow the corresponding EMTD input resolution, whereas dance videos are generated at $704\times1280$. We generate 121 frames at 24 FPS using 50 flow-matching sampling steps. Talking uses two-pass CFG with scale 5, while the two dance tasks use the three-pass CFG in Equation~\ref{eq:text-audio-cfg}, with text and audio scales of 5 and 4, respectively.

\paragraph{Evaluation Metrics.}
We evaluate reference fidelity, visual quality, audio--visual lip synchronization, motion--beat alignment, and human preference.

\textbf{Reference and visual quality.} Following VBench~\cite{huang2024vbench} and its I2V extension in VBench++~\cite{huang2024vbenchpp}, we report DINO-based Subject and DreamSim-based Background fidelity to the reference image, interpolation-based Motion Smoothness, LAION Aesthetic Quality, and MUSIQ Imaging Quality.

\textbf{Lip synchronization.} We use the official SyncNet evaluator~\cite{chung2017out} to report LSE-C and LSE-D. Higher LSE-C and lower LSE-D indicate stronger synchronization for the vocal-source role; the desired directions are reversed for the listener role, where synchronization with external vocals should be suppressed. Videos without a valid face track are excluded.

\textbf{Motion--beat alignment.} Kinematic beats are identified as strict local minima of a Gaussian-smoothed, scale-normalized motion-speed curve computed from 65 body, foot, and hand landmarks extracted by DWPose~\cite{yang2023effective}. BeatAlign~\cite{li2021aist} measures the Gaussian-weighted temporal proximity from each kinematic beat to its nearest musical beat, with $\sigma=0.05$ seconds. Because this motion-centered score can remain high when only a few well-timed movements are generated, we follow the coverage formulation of the Audio Beat Hit Score in TMD-Bench~\cite{yang2026tmdbench} and report an adapted Music-Beat Coverage Rate (MBCR). MBCR measures the fraction of musical beats covered by at least one kinematic beat within a 100-ms tolerance. The two metrics therefore characterize beat precision and beat coverage, respectively, and are averaged over the common pose-valid subset within each task.

\textbf{Human evaluation.} We conduct a blind pairwise study between SingDance and Wan-S2V. MusicInfuser is excluded because it does not condition on the reference image, making its identity apparent in a reference-conditioned comparison. We randomly sample 20 source-role song inputs from \textit{SingDance-50} and 20 instrumental inputs from \textit{Dance-100}. The instrumental restriction is applied only to the dancing-only subset, preventing Wan-S2V's default vocal articulation from revealing model identity. Fifteen participants evaluate all 40 matched-input, 24-FPS pairs with randomized and balanced A/B ordering, selecting the better result for body motion--music rhythm alignment and overall visual quality without ties. We report preference rates over 300 votes per task and criterion.

\begin{table}[!t]
\centering
{\small
\setlength{\tabcolsep}{2pt}
\begin{tabular*}{\columnwidth}{@{\extracolsep{\fill}}llcccc@{}}
\toprule
Role & Method & LSE-C & LSE-D & BeatAlign $\uparrow$ & MBCR $\uparrow$ \\
\midrule
Source & Wan-S2V & 4.2809 & 9.4462 & 0.2640 & 0.1983 \\
 & \textbf{SingDance} & \textbf{4.9672} & \textbf{8.9363} & \textbf{0.2723} & \textbf{0.2558} \\
\midrule
Listener & Wan-S2V & 4.1676 & 9.5386 & 0.2480 & 0.2065 \\
 & \textbf{SingDance} & \textbf{1.1768} & \textbf{12.4059} & \textbf{0.2704} & \textbf{0.2787} \\
\bottomrule
\end{tabular*}
}
\caption{Paired vocal-role switching on \textit{SingDance-50}. Source and listener correspond to Lip-Sync ON and OFF; LSE-C and LSE-D therefore have opposite desired directions.}
\label{tab:vocal-role-switch}
\end{table}

\begin{figure}[!t]
    \centering
    \includegraphics[width=.99\linewidth]{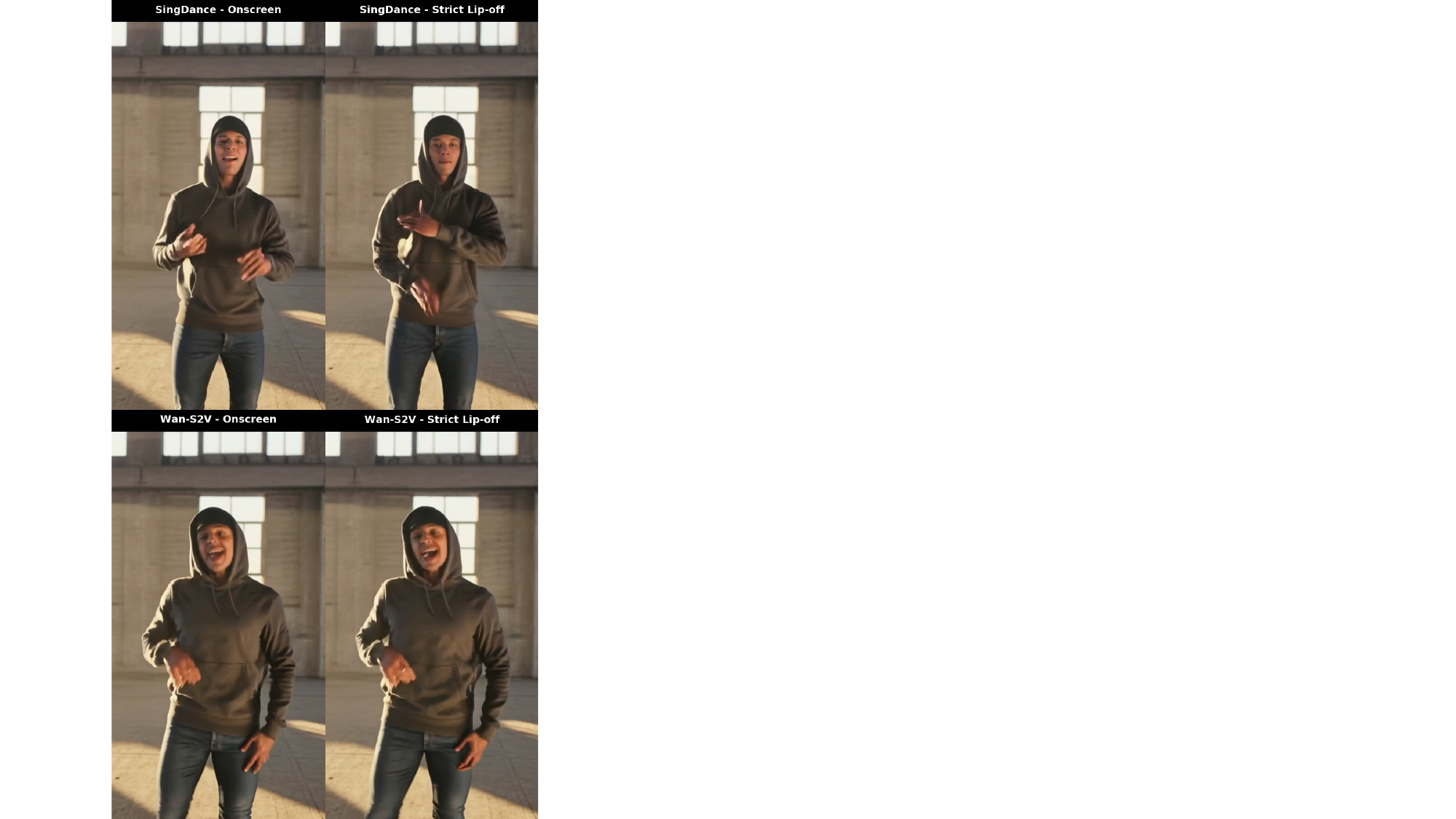}
    \caption{Qualitative paired vocal-role switching on two SingDance examples. Each Lip-on/Lip-off pair shares all role-independent inputs and inference settings; only the vocal role changes.}
    \label{fig:paired-vocal-role-switch}
\end{figure}

\subsection{Main Results}

Table~\ref{tab:main-dance} evaluates compositional zero-shot singing-and-dancing on \textit{SingDance-50} and directly supervised dancing-only generation on \textit{Dance-100}. It compares the dimensions shared by all applicable methods: reference fidelity, motion--beat alignment, visual quality, and human preference. We reserve vocal articulation for the paired analysis in the following subsection, where its desired role is explicitly controlled. The routed music token is separately ablated in Table~\ref{tab:music-ablation}.

\paragraph{Singing-and-dancing.}
Despite never observing paired singing-and-dancing videos during training, SingDance generates rhythmically coordinated performances while preserving the reference appearance and strong visual quality. Its advantage over speech-oriented generation demonstrates that music-specific conditioning complements text-guided choreography and vocal articulation. Human raters also prefer SingDance for rhythmic coordination and overall visual quality.

\paragraph{Dancing-only.}
On \textit{Dance-100}, SingDance continues to produce music-responsive body motion with strong reference preservation and visual fidelity. Its balanced performance across rhythmic coordination and generation quality shows that the unified model retains its directly supervised dancing-only capability while supporting compositional zero-shot singing-and-dancing. This advantage is also reflected in the human preference results.

\subsection{Paired Vocal-Role Switching}

Table~\ref{tab:main-dance} establishes generation quality and motion--beat alignment on the two task-specific test sets, but these metrics alone do not verify their defining behavioral distinction: whether the visible subject produces or merely hears the vocals. We therefore complement the main comparison with a paired intervention on \textit{SingDance-50}. The source rows reuse the corresponding singing-and-dancing outputs from Table~\ref{tab:main-dance}. For each listener counterpart, the reference image, song, role-independent prompt fields, checkpoint, sampling seed, and inference configuration remain fixed, while only the vocal-role state changes; its learned conditions and the prompt's \emph{vocal role} field are updated consistently. We apply the same textual switch to Wan-S2V as a prompt-only control, since it does not expose learned vocal-role conditions.

The prompt-only switch leaves Wan-S2V's articulation behavior largely unchanged, showing that text alone provides weak control over whether the visible subject owns the vocals. In contrast, SingDance consistently changes from vocal articulation to a listener role while retaining music-aligned body motion and stable visual quality. Figure~\ref{fig:paired-vocal-role-switch} shows two representative paired examples. Together with Table~\ref{tab:main-dance}, this intervention verifies that the same model can realize the defining behavioral difference between singing-and-dancing and dancing-only generation.

\begin{figure}[!t]
    \centering
    \includegraphics[width=\columnwidth]{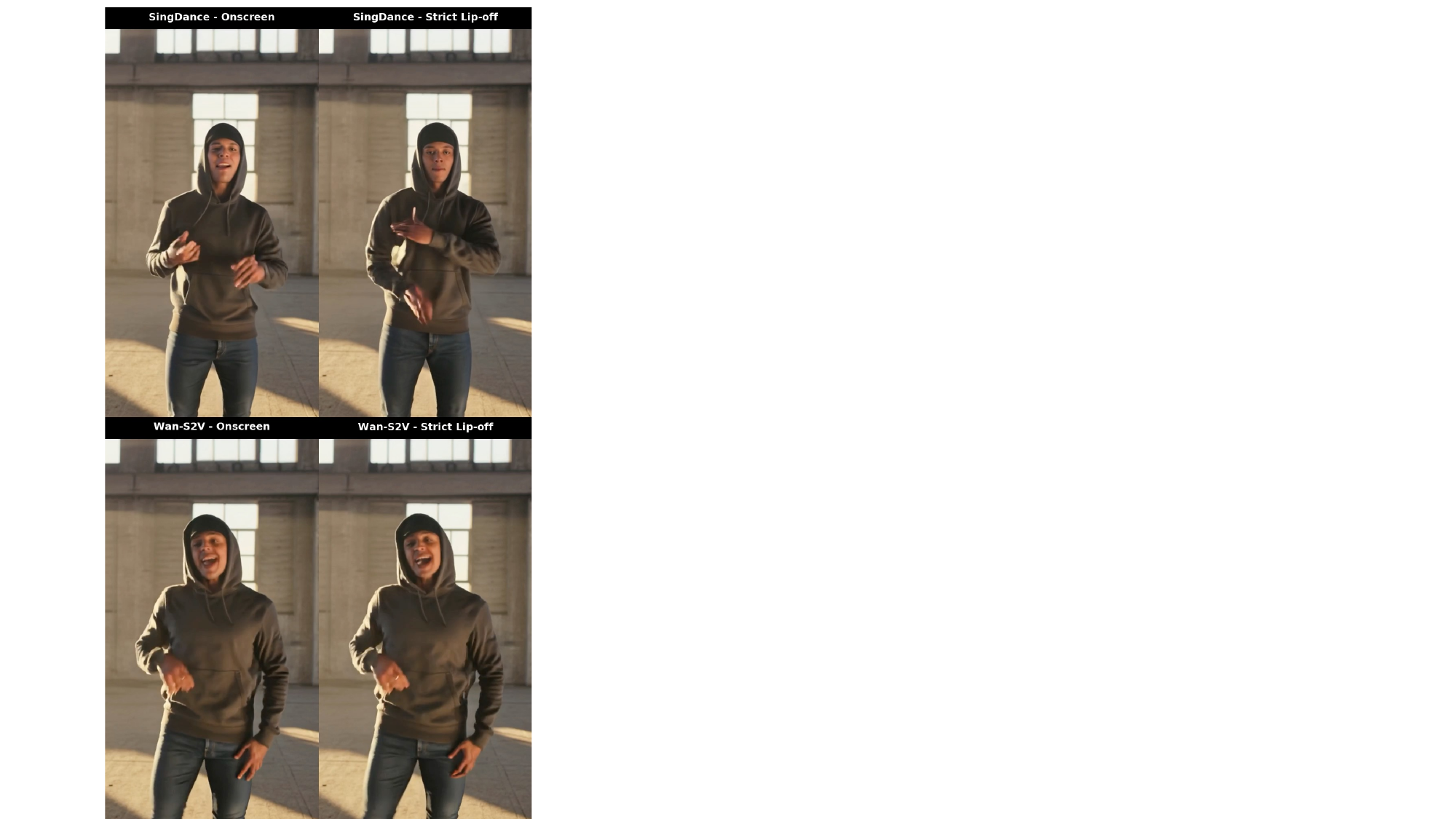}
    \caption{Qualitative vocal-role control with the same reference image and speech audio. SingDance produces synchronized articulation for the source role and listening/reacting behavior for the listener role; videos are provided in the supplementary material.}
    \label{fig:qualitative-vocal-articulation}
\end{figure}

\subsection{Vocal Articulation Capability}

Reliable source-role articulation is a prerequisite for compositional zero-shot singing-and-dancing. Table~\ref{tab:speech} compares SingDance with representative speech-driven systems, including Hallo3~\cite{cui2024hallo3}, Wan2.2-S2V~\cite{gao2025wans2v}, HY-Avatar~\cite{chen2025hunyuan}, and InfiniteTalk~\cite{yang2025infinitetalk}, on EMTD. Using roughly one-third of InfiniteTalk's generation-time parameters, SingDance achieves highly competitive lip synchronization together with the strongest aesthetic and imaging quality among the evaluated systems. This parameter-efficient articulation capability provides the foundation for compositional zero-shot singing-and-dancing, where synchronization must remain reliable alongside concurrent body motion rather than the relatively constrained motion of conventional talking clips.

\begin{table}[!t]
\centering
{\small
\setlength{\tabcolsep}{1.2pt}
\begin{tabular*}{\columnwidth}{@{\extracolsep{\fill}}lcccccc@{}}
\toprule
Method & Params. $\downarrow$ & LSE-C $\uparrow$ & LSE-D $\downarrow$ & Subj. $\uparrow$ & Aes. $\uparrow$ & Img. $\uparrow$ \\
\midrule
Hallo3 & 14.55 & 5.352 & 9.828 & 0.9746 & 0.511 & 0.677 \\
Wan-S2V & 16.30 & 6.999 & 8.292 & 0.9812 & \underline{0.580} & \underline{0.698} \\
HY-Avatar & \underline{12.99} & 7.210 & 8.494 & 0.9791 & 0.535 & 0.661 \\
InfiniteTalk & 18.88 & \textbf{8.535} & \textbf{7.108} & \textbf{0.9868} & 0.545 & 0.680 \\
\midrule
\textbf{SingDance} & \textbf{5.63} & \underline{7.995} & \underline{7.551} & \underline{0.9859} & \textbf{0.593} & \textbf{0.722} \\
\bottomrule
\end{tabular*}
}
\caption{Speech-driven vocal-articulation results on EMTD. Params. denotes generation-time parameters in billions; best and second-best results are bold and underlined.}
\label{tab:speech}
\end{table}

\subsection{Ablation Study}

Table~\ref{tab:music-ablation} isolates the contribution of the routed music token. Removing it consistently weakens motion--beat alignment on both test sets, while aesthetic and imaging quality remain broadly stable. These results show that the music pathway contributes effective rhythmic conditioning without compromising visual quality.

\begin{table}[!t]
\centering
{\small
\setlength{\tabcolsep}{1.2pt}
\begin{tabular*}{\columnwidth}{@{\extracolsep{\fill}}llcccc@{}}
\toprule
Test Set & Variant & BeatAlign $\uparrow$ & MBCR $\uparrow$ & Aes. $\uparrow$ & Img. $\uparrow$ \\
\midrule
\textit{SingDance-50} & w/o $M$ & 0.2462 & 0.2386 & \textbf{0.5864} & 0.6598 \\
 & Full & \textbf{0.2723} & \textbf{0.2558} & 0.5846 & \textbf{0.6688} \\
\midrule
\textit{Dance-100} & w/o $M$ & 0.2853 & 0.2457 & 0.5687 & 0.6734 \\
 & Full & \textbf{0.3002} & \textbf{0.2632} & \textbf{0.5718} & \textbf{0.6738} \\
\bottomrule
\end{tabular*}
}
\caption{Ablation of the routed music token on \textit{SingDance-50} and \textit{Dance-100}. Aes. and Img. denote Aesthetic and Imaging Quality; bold values indicate the better variant on each test set.}
\label{tab:music-ablation}
\end{table}

\section{Limitations}

SingDance targets short, single-person clips with a binary clip-level vocal role. It does not yet model within-clip role transitions, duets, multiple vocal roles, or automatic role attribution. Extending role-aware composition to longer, multi-person performances remains future work.

\section{Conclusion}

We introduced SingDance, a role-aware video generation framework that separates vocal content from the visible subject's role while coordinating text-guided action and music-conditioned body motion. Hard-compact routing and frame-wise joint audio injection compose speech, music, and role conditions within a shared video diffusion backbone. From asymmetric speaking/listening and dancing-only supervision, SingDance realizes compositional zero-shot singing-and-dancing in the held-out Song/Source configuration. Experiments validate music-aligned dance, paired vocal-role control, and parameter-efficient lip synchronization.

\bibliography{references}

\end{document}